\documentclass[conf]{new-aiaa}
\usepackage[utf8]{inputenc}

\usepackage{graphicx}
\usepackage{amsmath}
\usepackage[version=4]{mhchem}
\usepackage{siunitx}
\usepackage{longtable,tabularx}
\usepackage{booktabs}
\usepackage{adjustbox}
\usepackage{float}

\usepackage{nomencl}
\makenomenclature

\usepackage{xcolor}

\newcommand{\figref}[1]{Fig.~\ref{#1}}
\newcommand{\tabref}[1]{Table~\ref{#1}}

\newcommand{\secref}[1]{Section~\ref{#1}}
\newcommand{\obs}[0]{\boldsymbol{o}}

\newcommand{\state}[0]{\boldsymbol{x}}
\newcommand{\State}[0]{\mathcal{X}}
\newcommand{\action}[0]{\boldsymbol{u}}

\newcommand{\control}[0]{\boldsymbol{u}}
\newcommand{\controlmax}[0]{\control_{\rm max}}
\newcommand{\Control}[0]{\mathcal{U}}

\newcommand{\Reals}[0]{\mathbb{R}}

\newcommand{\radius}[0]{d}

\newcommand{\sunangle}[0]{\theta_{\rm Sun}}
\newcommand{\sunangledot}[0]{\dot{\theta}_{\rm Sun}}
\newcommand{\numpoints}[0]{n_p}

\newcommand{\admissibleset}[0]{\mathcal{C}_{\varphi}}
\newcommand{\safeset}[0]{\mathcal{C}_{S}}
\newcommand{\udes}[0]{\control_{\rm des}}
\newcommand{\uact}[0]{\control_{\rm act}}
\newcommand{\meanmotion}[0]{\eta}

\title{Space Processor Computation Time Analysis for Reinforcement Learning and Run Time Assurance Control Policies}

\author{
Kyle Dunlap\footnote{AI Software Developer, Intelligent Systems Division, AIAA Professional Member.}\textsuperscript{1} and 
Nathaniel Hamilton\footnote{AI Scientist, Intelligent Systems Division.}\textsuperscript{1}
}
\affil{Parallax Advanced Research, Beavercreek, OH, 45431, USA}

\author{
Francisco Viramontes\footnote{Research Scholar.}
}
\affil{COSMIAC, Albuquerque, NM, 87106, USA}

\author{
Derrek Landauer\footnote{Senior Computer Engineer, Space Systems Research.}
}
\affil{BlueHalo, Albuquerque, NM, 87123, USA}

\author{
Evan Kain\footnote{Principal Investigator for Space Computing, Space Electronics Technology (SET) Program.}
}
\affil{Air Force Research Laboratory, Kirtland Air Force Base, NM, 87117, USA}

\author{
Kerianne L. Hobbs\footnote{Safe Autonomy Lead, Autonomy Capability Team (ACT3), AIAA Associate Fellow.}
}
\affil{Air Force Research Laboratory, Wright-Patterson Air Force Base, OH, 45433, USA}
\begin{document}

\maketitle
\begingroup\renewcommand\thefootnote{1}
\begin{NoHyper}
\footnotetext{These authors contributed equally to this work.}
\end{NoHyper}
\endgroup

\begin{abstract}
As the number of spacecraft on orbit continues to grow, it is challenging for human operators to constantly monitor and plan for all missions. Autonomous control methods such as reinforcement learning (RL) have the power to solve complex tasks while reducing the need for constant operator intervention. By combining RL solutions with run time assurance (RTA), safety of these systems can be assured in real time. However, in order to use these algorithms on board a spacecraft, they must be able to run in real time on space grade processors, which are typically outdated and less capable than state-of-the-art equipment. In this paper, multiple RL-trained neural network controllers (NNCs) and RTA algorithms were tested on commercial-off-the-shelf (COTS) and radiation tolerant processors. The results show that all NNCs and most RTA algorithms can compute optimal and safe actions in well under 1 second with room for further optimization before deploying in the real world.
\end{abstract}

%%%%%%%%%%%%%%%%%%%%%%%%%%%%%%%%%%%%%%%%%%%%%%%%%%%%%%%%%%%%%%%%%%%%%%
% Nomenclature
%%%%%%%%%%%%%%%%%%%%%%%%%%%%%%%%%%%%%%%%%%%%%%%%%%%%%%%%%%%%%%%%%%%%%%
% \section{Nomenclature}

% \todo{nomenclature is a work in progress}

% \nomenclature{\(c\)}{Speed of light in a vacuum}
% \nomenclature{\(h\)}{Planck constant}

% \printnomenclature

% \section{Things to finish}
% \begin{itemize}
%     \item Introduction
%     \item What requirements must be met for space control? Not operating as fast as possible due to other required computation. Talk to Kerianne
%     \begin{itemize}
%         \item Contributions: Showing NNC + RTA can run on COTS
%         \item Showing NNC + RTA can run on rad-tolerant
%     \end{itemize}
%     \item Background
%     \begin{itemize}
%         \item Space Hardware/Challenges (COTS v RAD-TOL v RAD-HARD)
%         \item RL
%     \end{itemize}
%     \item Experimental setup
%     \begin{itemize}
%         \item Inspection task (RL and RTA setup)
%         \item Lab processors
%         \item Experiment descriptions (what we time)
%         \item metrics for comparison
%     \end{itemize}
%     \item Results
%     \begin{itemize}
%         \item Plots for important comparisons
%         \item Timing for COTS
%         \item Timing for rad-tolerant
%     \end{itemize}
%     \item Conclusion/Discussion
%     \begin{itemize}
%         \item Further optimizations to code
%     \end{itemize}
% \end{itemize}

%%%%%%%%%%%%%%%%%%%%%%%%%%%%%%%%%%%%%%%%%%%%%%%%%%%%%%%%%%%%%%%%%%%%%%
% Introduction
%%%%%%%%%%%%%%%%%%%%%%%%%%%%%%%%%%%%%%%%%%%%%%%%%%%%%%%%%%%%%%%%%%%%%%
\section{Introduction}
% \lettrine{T}{his} % supposed to use this for the first word apparently

% \todo{What's the problem?}
\lettrine{F}{or} \textit{On-orbit Servicing, Assembly, and Manufacturing} (OSAM) missions, docking and inspection tasks enable spacecraft operators to assess, plan for, and execute a wide variety of objectives. While these tasks are traditionally executed using classical control methods, this requires constant monitoring and adjustment by human operators, which becomes challenging or even impossible as the number of spacecraft increases. Therefore, the growing number of space assets and increasing complexity of operations makes the development of autonomous spacecraft operation critically important.  

% \todo{How can RL solve it?}
One promising solution for this problem is to utilize \textit{Deep Reinforcement Learning} (RL) to learn optimal control policies, in the form of Neural Network Controllers (NNCs), for executing tasks in increasingly complex scenarios. RL is a fast-growing field for developing high-performing autonomy with growing impact spurred by success in agents that learn to beat human experts in games like Go \cite{silver2016mastering}, Starcraft \cite{vinyals2019grandmaster}, and Gran Turismo \cite{wurman2022outracing}. RL has also demonstrated success reacting in real-time to changing mission objectives and environment uncertainty, which is crucial for spacecraft operations \cite{ravaioli2022safe, hamilton2022zero}. Additionally, RL has already demonstrated success in the space domain for inspecting illuminated (\cite{vanWijkAAS_23}) and uncooperative (\cite{brandonisio2021reinforcement}) space objects as well as collision-free docking \cite{oestreich2021autonomous, broida2019spacecraft, hovell2021deep, ravaioli2022safe, hamilton2023ablation}.

% \todo{What's the concern? RTA version (re-write to focus on this paper's concerns)}
Despite these successes using RL for complex, dynamic challenges, they are predominantly limited to virtual environments. RL agents trained in simulation to solve real-world problems can present unexpected and catastrophic behavior when deployed \cite{jang2019simulation, kadian2019we}. This is caused when the system enters scenarios that differ from the examples seen in training that produce a suboptimal control output that drives the system further from known scenarios that could be dangerous. Efforts in Safe RL (\cite{alshiekh2018safe, fulton2018safe, fisac2018general, zhao2020learning, hunt2021hscc, li2021safe, zhang2021safe, wagener2021safe, hamilton2022training, hamilton2023ablation, dunlap2023rta_rl}) have helped reduce these catastrophic events, but without formally verifying NNCs, there is no guarantee it will not happen. Therefore, to use high-performing NNCs in the real world, they must be bound by Run Time Assurance (RTA). 

% To protect systems, and speed up the training process, engineers can train their RL agents in simulation and then transfer the learned control policy to the corresponding real-world system. This process is a challenging problem referred to as the \textit{sim2real} challenge. Simulation environments abstract away a lot of the nuance and noise of the real world. As a result, RL policies trained in simulation can end up ``brittle''. When confronted with scenarios that differ from the examples seen in training they can fail to contextualize the situation and break. Thus, \textit{sim2real} transfers often have catastrophic results where the real world results are drastically different from simulation \cite{jang2019simulation, kadian2019we}. Therefore, in order to take advantage of the many benefits of training in simulation, it is critical to train more robust policies and/or formally verify their behavior before deploying on safety-critical, real-world systems. 

RTA is an online safety assurance technique for ensuring the safe operation of control systems by filtering the output of the primary control (e.g. a NNC), intervening as necessary to assure system safety \cite{hobbs2023runtime}. The RTA method focused on in this paper uses control barrier functions (CBFs) to assure safety, where an optimization algorithm is solved to obtain a safe control input as close as possible to the desired control input. Using CBFs allows RTA to enforce multiple safety constraints simultaneously.

% \todo{What's the concern? Computation version}
However, fixing the safety concerns using RTA introduces a new concern: computation time. Because developing space-grade hardware is a costly and lengthy process, the hardware tends to be outdated and has far less capability than state-of-the-art hardware \cite{lovelly2017comparative}. This has lead to concerns about running NNCs and RTA fast enough to control a spacecraft in real-time since using CBFs to assure safety requires continually solving an optimization algorithm in real time, which is a computationally intense. 
% \todo{What is this paper doing to help ease these concerns?}
This work seeks to assuage these concerns by measuring the execution times for NNCs and RTAs on a variety of space-grade hardware to identify expected timing requirements.

% \textbf{Our contributions.} 
The main contribution focuses on providing timing results from running NNCs and RTAs on space-grade hardware. Additionally this work presents comparisons that could lead to future improvements that could result in faster execution times. This work builds on previous research to train an RL agent to complete an inspection problem \cite{vanWijkAAS_23}, and developing RTA constraints for the same task \cite{dunlap2023run}. The results from this work demonstrate that, even without further optimization, NNCs and RTA can produce a safe and optimal control value in under 20ms\footnote{This is assuming the RTA is explicit, continuous ASIF.} on COTS and radiation tolerant processors.

%%%%%%%%%%%%%%%%%%%%%%%%%%%%%%%%%%%%%%%%%%%%%%%%%%%%%%%%%%%%%%%%%%%%%%
% Background
%%%%%%%%%%%%%%%%%%%%%%%%%%%%%%%%%%%%%%%%%%%%%%%%%%%%%%%%%%%%%%%%%%%%%%
\section{Background}

This section provides introductions to reinforcement learning, run time assurance, and the types of space grade hardware used in this work.

\subsection{Reinforcement Learning}

\emph{Reinforcement Learning} is a form of machine learning in which an agent acts in an environment, learns through experience, and improves the performance of the learned behaviour function, i.e. policy, based on the observed reward. The environment can be comprised of any dynamical system, from Atari simulations (\cite{hamilton2020sonic, alshiekh2018safe}) to complex robotics scenarios (\cite{brockman2016openai, fisac2018general, henderson2018deep, mania2018simple, jang2019simulation, bernini2021few}).

In \emph{Deep Reinforcement Learning} (RL) the policy is approximated using a deep neural network, which becomes the \emph{Neural Network Controller} (NNC) after training is completed. This work focuses on NNCs trained using the Proximal Policy Optimization (PPO) algorithm. PPO was first introduced in \cite{schulman2017proximal}. The  ``proximal'' in PPO refers to how the algorithm focuses on iteratively improving the policy in small increments to prevent against making large changes in the policy that lead to drops in performance. PPO was selected because of its widespread use in a variety of RL tasks and previous demonstrated success in the space domain \cite{broida2019spacecraft, oestreich2021autonomous, hamilton2023ablation, dunlap2023rta_rl, vanWijkAAS_23}.

% \todo{Feel like I hit all the important parts, but think I'm missing anything crucial?}

\subsection{Run Time Assurance}

\begin{figure}[htbp]
    \centering
    \includegraphics[width=.8\columnwidth]{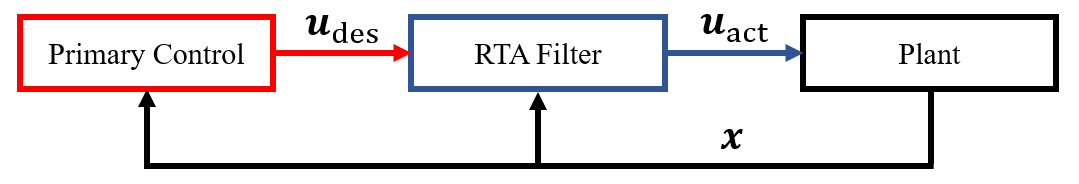}
    \caption{Feedback control system with RTA filter. Components with low safety confidence are outlined in red, and components with high safety confidence are outlined in blue.}
    \label{fig:RTA_Filter}
\end{figure}

Run Time Assurance (RTA) is an online safety assurance technique that filters potentially unsafe inputs from a primary controller and intervenes as necessary to assure system safety \cite{hobbs2023runtime}. The primary controller can come in many different forms, such as the NNC trained with RL, a classical control technique, or even a human operator. It is decoupled from the RTA filter, and can completely focus on completing the task while the RTA filter focuses on assuring safety. A feedback control system with RTA is shown in \figref{fig:RTA_Filter}. The primary controller receives the state $\state$ and computes a desired control input $\udes$ to pass to the RTA filter. The RTA filter evaluates $\udes$ based on the current state and modifies it as necessary to produce a safe control input $\uact$ to pass to the plant.

For a continuous time control affine dynamical system defined as,
\begin{equation} \label{eq:fxgu}
   \dot{\state} = f(\state) + g(\state)\control,
\end{equation}
where $f:\State \rightarrow \Reals^n$ and $g:\State \rightarrow \Reals^{n \times m}$ are locally Lipschitz continuous functions. $\state \in \State \subseteq \Reals^n$ denotes the state vector, $\control \in \Control \subseteq \Reals^m$ denotes the control vector, $\State$ is the set of all possible state values, and $\Control$ is the admissible control set.

To define safety of this system at the current time, consider the superlevel set of a continuously differentiable function $\varphi:\State \rightarrow \Reals$ defined as $\admissibleset = \{\state \in \State : \varphi(\state) \ge 0\}$, where $\admissibleset$ is referred to as the allowable set. The system will be safe for all time if it is in a forward invariant subset of $\admissibleset$ known as the safe set $\safeset \subseteq \admissibleset \subset \Reals^n$, where for every $\state \in \safeset$, $\state(t) \in \safeset, \, \forall t \geq 0$. A continuously differentiable function $h:\State \rightarrow \Reals$ is used to define $\safeset$, where $\safeset = \{\state \in \State : h(\state) \ge 0\}$.
% To define safety of this system, consider the superlevel set of a continuously differentiable function $h:\State \rightarrow \Reals$ defined as $\safeset = \{\state \in \State : h(\state) \ge 0\}$, where $\safeset$ is referred to as the safe set. The system will be safe if $\safeset$ is forward invariant (it cannot leave $\safeset$), where for every $\state \in \safeset$, $\state(t) \in \safeset, \, \forall t \geq 0$.
To enforce forward invariance of $\safeset$, Nagumo's condition \cite{nagumo1942lage} is used, where the boundary of $\safeset$ is examined to ensure $\dot{h}(\state) \geq 0$. To relax this condition away from the boundary, a continuous function $\alpha : \Reals \rightarrow \Reals$ is introduced, where $\alpha$ is an extended class $\mathcal{K}$ function that is strictly increasing and has the property $\alpha(0) = 0$.
Safety can be enforced using Control Barrier Functions (CBFs) \cite{ames2019control}, where $h$ is a CBF if there exists an extended class $\mathcal{K}$ function $\alpha$ and control $\control \in \Control$ such that,
\begin{equation}\label{eq:cbf_condition}
    \sup_{\control \in \Control} [L_fh(\state) + L_gh(\state) \control ] \geq - \alpha(h(\state)), ~ \forall \state \in \safeset,
\end{equation}
where $L_f$ and $L_g$ are Lie derivatives of $h$ along $f$ and $g$ respectively.

To enforce safety inside the control loop, this paper uses the RTA technique known as the Active Set Invariance Filter (ASIF) \cite{ASIF_2018}. ASIF is a first-order optimization algorithm that is minimally invasive towards the primary controller. It is formulated as a Quadratic Program (QP), where the objective function is the $l^2$ norm difference between $\udes$ and $\uact$. CBFs are defined as inequality constraints, known as barrier constraints, and are used to enforce safety where,
\begin{equation}
    BC(\state, \control) := L_f h(\state) + L_g h(\state) \control + \alpha(h(\state)) \geq 0.
\end{equation}
For $N$ barrier constraints, the ASIF algorithm is defined as,

\begin{samepage}
\noindent \rule{1\columnwidth}{0.7pt}
\noindent \textbf{Active Set Invariance Filter}
\begin{equation}
\begin{gathered}
\control_{\rm act}(\state, \control_{\rm des})= \underset{\control \in \Control}{\text{argmin}} \left\Vert \control_{\rm des}-\control\right\Vert\\
\text{s.t.} \mkern9mu BC(\state, \control) \geq 0, \mkern9mu \forall i \in \{1,...,N\}
\end{gathered}\label{eq:optimization}
\end{equation}
\noindent \rule[7pt]{1\columnwidth}{0.7pt}
\end{samepage}

For the dynamical system in \eqref{eq:fxgu}, the barrier constraints can be defined explicitly as,
\begin{equation} \label{eq:exp_BC}
    BC_i(\state,\control) := \nabla h_i(\state) (f(\state) + g(\state)\control) + \alpha_i(h_i(\state)).
\end{equation}
In this case, safety is defined offline to ensure the CBFs render $\safeset$ forward invariant. This RTA technique will be referred to as explicit ASIF (eASIF).

Safety can also be defined implicitly online, where closed loop trajectories under a predetermined backup control law $\control_{\rm b}$ are continuously computed to ensure $\safeset$ can be rendered forward invariant.
% In this case, safety at the current time is defined by the superlevel set of a continuously differentiable function $\varphi:\State \rightarrow \Reals$, defined as $\admissibleset = \{\state \in \State : \varphi(\state) \ge 0\}$, where $\admissibleset$ is referred to as the allowable set and $\safeset \subseteq \admissibleset \subset \Reals^n$. 
In this case, the barrier constraint is based on $\admissibleset$, and is defined implicitly as,
\begin{equation} \label{eq:imp_BC}
    BC_i(\state,\control) := \nabla \varphi_i(\phi^{\control_{\rm b}}_j) D(\phi^{\control_{\rm b}}_j) [f(\state) + g(\state)\control] + \alpha_i(\varphi_i(\phi^{\control_{\rm b}}_j) ),
\end{equation}
where $\phi^{\control_{\rm b}}_j$ defines a prediction of the state $\state$ at the $j^{th}$ discrete time interval $\forall t \in [0,\infty)$ along the trajectory computed under $\control_{\rm b}$, and $D(\phi^{\control_{\rm b}}_j)$ is computed by integrating a sensitivity matrix along this trajectory \cite{chen2021backup}. This RTA technique will be referred to as implicit ASIF (iASIF). Note that for practical implementation, the trajectories are limited to a finite time horizon $T$.

Finally, safety can also be computed in discrete time, which can be useful when the RTA filter must be run at a lower frequency. For a discrete time dynamical system defined as,
\begin{equation} \label{eq: discrete_dynamics}
    \state_{t + \Delta t} = f(\state_t, \control_t),
\end{equation}
a continuously differentiable function $h : \State \rightarrow \Reals$ is a discrete CBF (DCBF) \cite{DCBF_2023} if there exists an extended class $\mathcal{K}$ function $\alpha$ satisfying $\alpha(r) < r, \forall r > 0,$ and $\control \in \Control$ such that,
\begin{equation} \label{eq:discreteInvariance}
    \Delta h(\state_t, \control_t) + \alpha(h(\state_t)) \ge 0, \mkern9mu \forall \state_t \in \safeset,
\end{equation}
where,
\begin{equation}
    \Delta h(\state_t, \control_t) = \frac{h(\state_{t + \Delta t}) - h(\state_t)}{\Delta t}.
\end{equation}
The discrete barrier constraint is therefore defined as,
\begin{equation} \label{eq:Discrete-BC}
    BC_i(\state_t,\control_t) := \Delta h_i(\state_t , \control_t) + \alpha_{i}(h_i(\state_t)) \ge 0.
\end{equation}
Notably, due to the computation of $\state_{t + \Delta t}$, these constraints often become nonlinear. Therefore, a nonlinear solver is used to solve the ASIF algorithm rather than a QP. In general, nonlinear solvers are incomplete and are much slower than QPs. This RTA technique will be referred to as discrete ASIF (dASIF).

\subsection{Space Hardware}

Due to radiation hazards that exist in space such as galactic cosmic rays, solar particle events, and trapped radiation in the Van Allen belts, processors on board spacecraft must be specially designed to withstand the space environment \cite{lovelly2017comparative}. Without this, radiation can cause issues such as data corruption or a malfunction of the processor. However, because developing space grade hardware is often a costly and lengthy process, this hardware tends to be outdated and have far less capability than state of the art hardware. This makes it challenging to run modern control algorithms that are designed to be run in real time.

Space grade hardware can be divided into three main categories: \textit{Commercial Off The Shelf} (COTS), radiation tolerant, and radiation hardened. COTS hardware is not specifically designed to withstand radiation, and as a result it is generally used for shorter term missions in \textit{Low Earth Orbit} (LEO) and not for missions in \textit{Geostationary Earth Orbit} (GEO). Radiation tolerant hardware is able to withstand some radiation, and is generally used for medium term missions in LEO or short term missions in GEO. Radiation hardened hardware can withstand a significant amount of radiation, and can be used for long term missions in LEO or GEO. While ideally all processors used in space would be radiation hardened, the designer must balance cost, availability, and functionality when choosing processors for their mission.

%%%%%%%%%%%%%%%%%%%%%%%%%%%%%%%%%%%%%%%%%%%%%%%%%%%%%%%%%%%%%%%%%%%%%%
% Experimental Setup
%%%%%%%%%%%%%%%%%%%%%%%%%%%%%%%%%%%%%%%%%%%%%%%%%%%%%%%%%%%%%%%%%%%%%%
\section{Experimental Setup}

This section introduces the spacecraft inspection task, the NNC architecture and RTA constraints to be tested, the specific processors used for experimentation, and a description of all experimental configurations.

\subsection{Inspection Task}
\begin{figure}[t]
    \centering
    \includegraphics[width=.5\columnwidth]{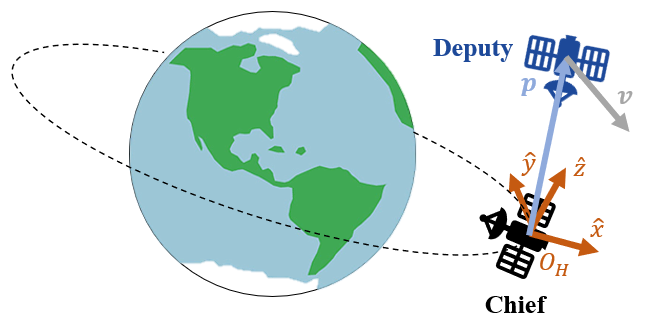}
    \caption{Deputy spacecraft in relation to a chief spacecraft in Hill's Frame.}
    \label{fig:Hills}
\end{figure}

In this work, the NNCs and RTA are designed for completing the inspection task introduced in \cite{vanWijkAAS_23}. The inspection task considers an active ``deputy'' spacecraft attempting to inspect a passive ``chief'' spacecraft \cite{vanWijkAAS_23}. The analysis takes place in Hill's frame \cite{hill1878researches}, shown in \figref{fig:Hills}, where the chief is in a circular orbit around the Earth, and the agent controls the deputy. The chief is represented by a sphere of 99 uniformly distributed points, where points can only be inspected if they are within the deputy's field of view and are illuminated by the Sun.

\subsubsection{Dynamics}

The dynamics for the inspection task are modeled using the Clohessy-Wiltshire equations \cite{clohessy1960terminal} in Hill's frame, where the origin $\mathcal{O}_H$ is located at the chief's center of mass, the unit vector $\hat{x}$ points away from the center of the Earth, $\hat{y}$ points in the direction of motion of the chief, and $\hat{z}$ is normal to $\hat{x}$ and $\hat{y}$.
The linearized relative motion dynamics between the deputy and chief are defined as, 
\begin{equation} \label{eq: system dynamics}
    \dot{\state} = A {\state} + B\action,
\end{equation}
where $\state$ is the state vector $\state=[x,y,z,\dot{x},\dot{y},\dot{z}]^T \in \Reals^6$, $\action$ is the control vector, $\action= [F_x,F_y,F_z]^T \in [-\controlmax, \controlmax]^3$, and,
\begin{align}
\centering
    A = 
\begin{bmatrix} 
0 & 0 & 0 & 1 & 0 & 0 \\
0 & 0 & 0 & 0 & 1 & 0 \\
0 & 0 & 0 & 0 & 0 & 1 \\
3\meanmotion^2 & 0 & 0 & 0 & 2\meanmotion & 0 \\
0 & 0 & 0 & -2\meanmotion & 0 & 0 \\
0 & 0 & -\meanmotion^2 & 0 & 0 & 0 \\
\end{bmatrix}, 
    B = 
\begin{bmatrix} 
 0 & 0 & 0 \\
 0 & 0 & 0 \\
 0 & 0 & 0 \\
\frac{1}{m} & 0 & 0 \\
0 & \frac{1}{m} & 0 \\
0 & 0 & \frac{1}{m} \\
\end{bmatrix}.
\end{align}
Here, $\meanmotion=0.001027$rad/s is the mean motion of the chief's orbit, $m=12$kg is the mass of the deputy, $F$ is the force exerted by the deputy's thrusters along each axis, and $\controlmax=1$N is the maximum force. The attitude of the deputy is not modeled, and it is assumed that it is always pointing towards the chief. 

The Sun in assumed to remain in a constant position in relation to the Earth, where it appears to rotate in Hill's frame. For this task, the Sun rotates in the $\hat{x}-\hat{y}$ plane in Hill's frame, where the unit vector pointing from the center of the chief spacecraft to the Sun, $\hat{r}_{S}$, is defined as,
\begin{equation}
    \hat{r}_{S} = [\cos{\sunangle}, \sin{\sunangle}, 0],
\end{equation}
where $\sunangledot = -\meanmotion$. It is also assumed that nothing blocks the chief from being illuminated.

\subsubsection{RL Environment Setup}\label{sec:RL_setup}

The RL environment is partially observable, where sensors are used to consolidate full state information into components of the observation space. In addition to the state $[x,y,z,\dot{x},\dot{y},\dot{z}]$ and sun angle $\sunangle$ as defined above, the agent receives two additional observations related to the inspection points: the total number of points that have been inspected during the episode $\numpoints$, and a unit vector pointing towards the nearest cluster of uninspected points $\hat{r}_{UPS}=[x_{UPS}, y_{UPS}, z_{UPS}]$, as determined by k-means clustering. Two observation spaces are considered for this analysis, where the first is referred to as ``no sensors'' and is defined as,
\begin{equation}
    \obs_{\rm no} = [x, y, z, \dot{x}, \dot{y}, \dot{z}].
\end{equation}
The second observation space is referred to as ``all sensors'' and is defined as,
\begin{equation}
    \obs_{\rm all} = [x, y, z, \dot{x}, \dot{y}, \dot{z}, \numpoints, \sunangle, x_{UPS}, y_{UPS}, z_{UPS}].
\end{equation}

\subsubsection{Safety Constraints}

While many safety constraints could be developed for the inspection task, the following constraints are enforced for this analysis. First, the deputy shall not collide with the chief. The CBF for this constraint is defined as,
\begin{equation}
    h_1(\state) := \sqrt{2 a_{\rm max} [\Vert \boldsymbol{p} \Vert_2 - (\radius_{\rm d}+\radius_{\rm c})]} + \boldsymbol{v}_{{prc}} \geq 0,
\end{equation}
where $a_{\rm max}$ is the maximum acceleration of the deputy, $\boldsymbol{p}$ is the deputy's position, $\radius_{\rm d}$ and $\radius_{\rm c}$ are the collision radii of the deputy and chief respectively, and $\boldsymbol{v}_{{prc}}$ is the projection of the deputy's velocity in the direction of the chief. This constraint ensures that the deputy can slow down to avoid a collision.

Second, the deputy shall remain in proximity with the chief so that it stays on task. This is treated as a keep in zone, and the CBF for this constraint is defined as,
\begin{equation}
    h_2(\state) := \sqrt{2 a_{\rm max} (\radius_{\rm max} - \Vert \boldsymbol{p} \Vert_2)} - \boldsymbol{v}_{{prr}} \geq 0,
\end{equation}
where $\radius_{\rm max}$ is the maximum relative distance from the chief, and in this case $\boldsymbol{v}_{{prr}}$ is the projection of the deputy's velocity in the direction of the keep in zone.

Third, the deputy shall slow down as it approaches the chief, which reduces the risk of a potential collision in the event of a fault. The CBF for this constraint is defined as,
\begin{equation}
    h_3(\state) := \nu_0 + \nu_1\Vert \boldsymbol{p} \Vert_2 - \Vert \boldsymbol{v} \Vert_2 \geq 0,
\end{equation}
where $\nu_0$ is a minimum allowable speed at the origin, $\nu_1$ is a constant rate at which $\boldsymbol{p}$ shall decrease, and $\boldsymbol{v}$ is the deputy's velocity.

Finally, the deputy shall not maneuver aggressively with high velocities. This is defined as three separate CBFs,
\begin{equation}
    h_4(\state) := v_{\rm max}^2 - \dot{x}^2\geq 0, \quad h_5(\state) := v_{\rm max}^2 - \dot{y}^2\geq 0, \quad h_6(\state) := v_{\rm max}^2 - \dot{z}^2\geq 0,
\end{equation}
where $v_{\rm max}$ is the maximum allowable velocity. Further details for all constraints can be found in \cite{dunlap2023run}.

The safe set is then defined as $\safeset = \{\state \in \State : h_i(\state) \ge 0\, , \, \forall i \in \{1, ..., 6\} \}$. The CBFs are enforced simultaneously by the ASIF algorithm. For simplicity and equal comparison, the same CBFs are used for explicit, implicit, and discrete ASIF.

\subsection{Neural Network Controllers}

\begin{figure}
    \centering
    \includegraphics[width=\columnwidth]{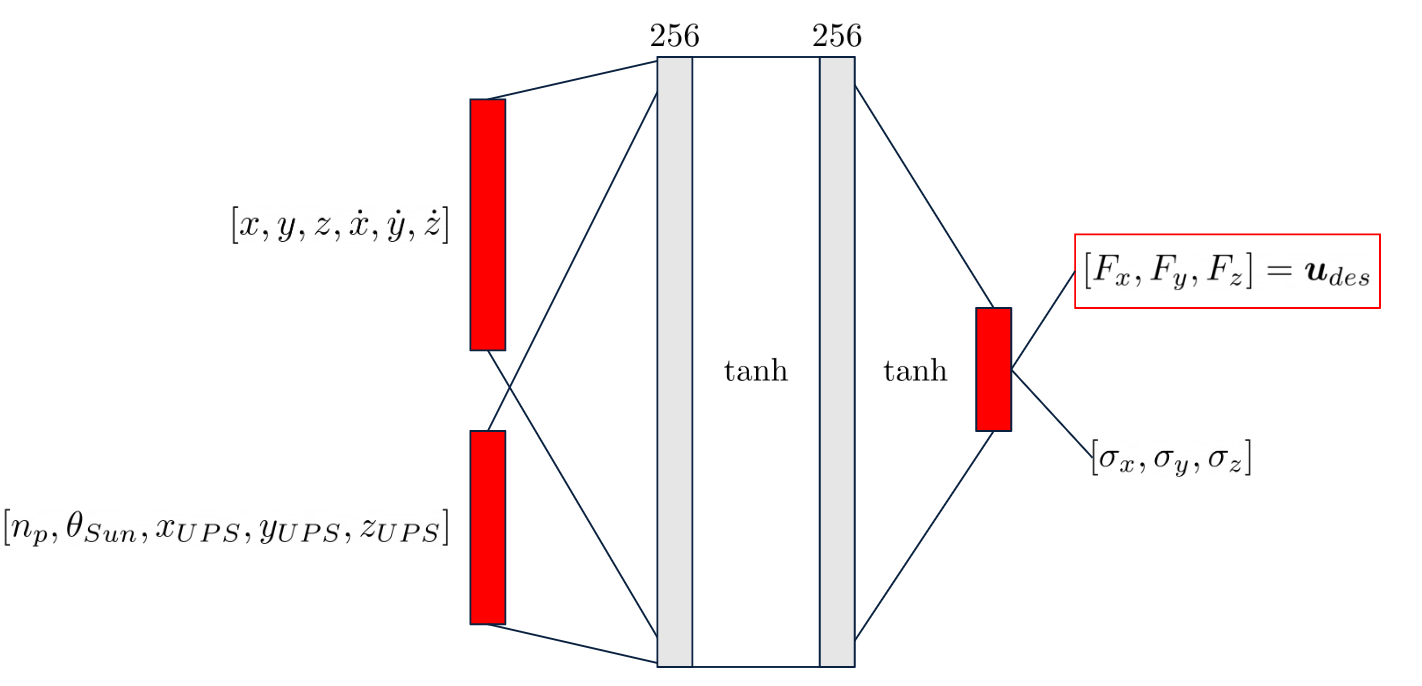}
    \caption{The NNC architecture for the ``all sensors'' variation of the inspection task. %The inputs to the NNC are the \todo{should these be red or a different color?} red blocks on the left. The output used for controlling the deputy spacecraft is the portion on the right in the red box. Part of the output is ignored because it is only used during the RL training. Note: the ``no senosrs'' variation is the same architecture but with the bottom input block removed.
    }
    \label{fig:nnc_architecture}
\end{figure}

This work focuses on testing two NNCs trained to solve the two variations of the inspection task described in \secref{sec:RL_setup}. The hyperparameters used to train the NNCs using the PPO algorithm for these experiments are listed in \tabref{tab:Hyperparameters}. The NNC architecture for the ``all sensors'' variation is shown in \figref{fig:nnc_architecture}. The ``no sensors'' variation has the same architecture, but with the bottom input block removed.

The NNC architecture is a fully-connected Multi-Layer Perceptron (MLP) network with two hidden layers with 256 nodes each. The hidden layers have a $\tanh$ activation function\footnote{This is the common/default architecture used for PPO.}. The NNC has either 6 inputs or 11 inputs, depending on the observation space used. The output layer consists of 6 nodes, which represent the mean and variance for each control value in $\action$ for a stochastic policy used during training. In these experiments, the output is always the mean value as show by the red box in \figref{fig:nnc_architecture}. The inputs are normalized such that typical values fall within the range $[-1, 1]$, where $[x, y, z]$ are each divided by $100$, $[\dot{x}, \dot{y}, \dot{z}]$ are each multiplied by $2$, and $\numpoints$ is divided by 100. Further details of the RL environment and training process can be found in \cite{vanWijkAAS_23}. 

\begin{table}[htb]
    \centering
    \caption{PPO Training Hyperparameters}
    \begin{tabular}{lc} \hline
        Parameter                       & Value \\ \hline
        Number of SGD iterations        & 30 \\ 
        Discount factor $\gamma$        & 0.99 \\
        GAE-$\lambda$                   & 0.928544 \\
        Max episode length              & 1223 \\
        Rollout fragment length         & 1500 \\
        Train batch size                & 1500 \\
        SGD minibatch size              & 1500 \\
        Total timesteps                 & $5 \times 10^6$ \\
        Learning rate                   & $5 \times 10^{-5}$ \\
        KL initial coefficient          & 0.2 \\
        KL target value                 & 0.01 \\
        Value function loss coefficient & 1.0 \\
        Entropy coefficient             & 0.0 \\
        Clip parameter                  & 0.3 \\
        Value function clip parameter   & 10.0 \\
        \hline
    \end{tabular}
    \label{tab:Hyperparameters}
\end{table}

%%%%%%%%%%%%%%%%%%%%%%%%%%%%%%%%%%%%%%%%%%%%%%%%%%%%%%%%%%%%%%%%%%%%%%
% Lab Processors
%%%%%%%%%%%%%%%%%%%%%%%%%%%%%%%%%%%%%%%%%%%%%%%%%%%%%%%%%%%%%%%%%%%%%%
\subsection{Space Processors}

% This section details the specific processors that were used for testing. \nph{I feel like this should be a part of exp setup and should include a table that compares the different boards and their specs}

The processors used for testing are listed in \tabref{tab:processors}, which were provided by the Spacecraft Processing Architectures and Computing Environment Research (SPACER) laboratory. For this analysis, three COTS processors and one radiation tolerant processor were tested. For all tests, only the CPU was used to estimate worst case performance in the real world. Additionally, all tests recorded the time from receiving the inputs to publishing the output.

\begin{table}[htb]
    \centering
    \caption{Processor Specifications}
    \resizebox{\linewidth}{!}{
    \label{tab:processors}
    \begin{tabular}{lrrrrrc}
    \toprule
    Board             & Radiation    & Cores & CPU Max Frequency & Memory Speed & Power          & GPU/FPGA                               \\
    \midrule
    Jetson AGX Orin   & COTS         & 12    & 2.2~GHz           & 204.8~GB/s   & 15 - 60~W      & \checkmark / X                         \\
    Jetson Orin Nano  & COTS         & 6     & 1.5~GHz           & 68~GB/s      & 7 - 15~W       & \checkmark / X                         \\
    Jetson AGX Xavier & COTS         & 8     & 2.2~GHz           & 136.5~GB/s   & 10 - 30~W      & \checkmark / X                         \\
    Unibap iX10       & Rad-Tolerant & 4     & 3.6~GHz           & 2.34~GB/s    & <40~W          & \checkmark / \checkmark                 \\
    \bottomrule
    \end{tabular}
    }
\end{table}

\subsubsection{COTS}

For all tests on the COTS processors, Python was used to run the NNC and RTA. The NNCs were trained using the Air Force Research Laboratory's Core Reinforcement Learning library (CoRL) \cite{merrick2023corl} and the RTA was originally developed in the Safe Autonomy Run Time Assurance Framework \cite{ravaioli2023universal}, which are both written in Python and allowed for a simple transition to the processors.
After training the NNCs, ONNX Runtime \cite{onnxruntime} was used to convert them into simple models that could be run on the processors. For the RTA, Quadprog \cite{quadprog} was used for the ASIF QP, and SciPy's SLSQP algorithm \cite{2020SciPy-NMeth} was used as the nonlinear solver for dASIF. NumPy \cite{harris2020array} was used for all linear algebra. ROS2 \cite{ros2} was used to run the algorithms on the processors, which represents a realistic framework for passing messages in data in a similar manner to what would be used on board a spacecraft. 
% The algorithms were timed from receiving the input state to publishing the output control to measure the entire time needed to run the algorithm in practice. \todo{all timing results measure the time...}

\subsubsection{Radiation Tolerant}

For all tests on the Unibap iX10, C++ was used to run the NNC and RTA. While it may have been possible to run Python on this processor, C and C++ are much more commonly used on space grade hardware. Therefore, an effort was made to ensure the algorithms tested were of this format. ONNX Runtime was used to run the NNC. For the RTA, QuadProg++ \cite{QuadProg++} was used for the QP, and the NLopt COBYLA agorithm \cite{NLopt} was used as the nonlinear solver. Eigen \cite{eigenweb} was used for linear algebra. For these tests, simple scripts were made to pass data and record the computation time.
% , where the time was recorded from receiving the inputs to publishing the output. \todo{all timing results measure the time...}

%%%%%%%%%%%%%%%%%%%%%%%%%%%%%%%%%%%%%%%%%%%%%%%%%%%%%%%%%%%%%%%%%%%%%%
% Experiments
%%%%%%%%%%%%%%%%%%%%%%%%%%%%%%%%%%%%%%%%%%%%%%%%%%%%%%%%%%%%%%%%%%%%%%
\subsection{Experiments}

Many different configurations were tested on the COTS and radiation tolerant processors. At a high level, these configurations can be divided intro three groups: NNC only, RTA only, and NNC combined with RTA. Each configuration was tested with 1,000 test cases, each corresponding to a different state. These states were divided into two groups: ``safe'' states that are guaranteed to be within $\safeset$, and ``not safe'' states that may or may not be in $\safeset$.
For the configurations that include an NNC, the NNC with all sensors (as defined in \secref{sec:RL_setup}) was tested on the COTS processors, and NNCs with either no sensors or all sensors were tested on the iX10. For the RTA only configurations, a set of 1,000 random actions were used to represent the primary controller. 

For all RTA configurations, each different RTA filter was considered. First, eASIF was tested, which represents the baseline configuration. Next, iASIF was tested with two different timesteps for the computed backup trajectory: 1~s and 10~s. The total trajectory length $T=20$~s in each case, so this results in either 20 or 2 computed trajectory steps. Finally, dASIF was also tested with the same two timesteps, which in this case correspond to when the next state is computed. For only the tests on the iX10, two different tolerances were also considered: 1e-3 and 1e-4. These tolerances are the relative and absolute tolerances of the optimization algorithm, and they control how accurate the solution is which directly affects computation time.

% \todo{Describe all configurations and setup} \nph{again, might fit better in exp setup}

% \todo{re-write for our experiments. This has the more accurate terms for what we are presenting}
% For benchmarking purposes, we recorded the mean execution-times (Mean ET) of our real-time reachability algorithm, as well as the average number of iterations utilized in constructing the reachable set (Mean Iters). While typically a discussion of upper bounds on execution times involves a discussion of the \emph{Worst-Case Execution Time} (WCET), we instead report the Mean ET. In general, the WCET is unknown or difficult to derive without the use of static analysis proofs \cite{wilhelm2008worst}. Since our safety regime relies on ROS, which is highly dynamic and distributed, it is prohibitively difficult to perform an exhaustive exploration of the space of all execution times and thus derive the WCET.  However, we provide a rough proxy of the WCET by reporting the \emph{Maximal Observed Execution Times} (MOET) \cite{wilhelm2008worst}.

For benchmarking purposes, the execution time for each test case was recorded. The main metric presented is the \textit{InterQuartile Mean} (IQM) of the execution time, which discards the highest and lowest 25\% of the recorded data and calculates the mean score on the remaining middle 50\% of data. The IQM is more robust to outliers, and is a better representation of the expected execution time \cite{agarwal2021deep}. The worst case execution time is also of interest, but it is often unknown or difficult to derive \cite{wilhelm2008worst}. Therefore, the \emph{Maximal Observed Execution Time} (MOET) is presented instead. Additional metrics are provided in \secref{sec:results} for completeness. Also note that ONNX Runtime uses Just In Time (JIT) compilation, meaning the model is compiled during the first time it is run. This time is separated into a separate column for \tabref{tab:nnc_ix10} and \tabref{tab:nnc_rta_ix10}.

%%%%%%%%%%%%%%%%%%%%%%%%%%%%%%%%%%%%%%%%%%%%%%%%%%%%%%%%%%%%%%%%%%%%%%
% Results & Discussion
%%%%%%%%%%%%%%%%%%%%%%%%%%%%%%%%%%%%%%%%%%%%%%%%%%%%%%%%%%%%%%%%%%%%%%
\section{Results} \label{sec:results}

This section provides all the timing results collected from the COTS and radiation tolerant hardware. A discussion of trends found in selected results are presented in \secref{sec:discussion}.

\subsection{COTS Timing Results}

The full timing results from running the experiments on COTS hardware are shown in the following tables. These results come from measuring the execution time for 1,000 test cases for each row. \tabref{tab:nnc_cots} shows the timing result executing only the NNC on all COTS processors. \tabref{tab:rta_cots} shows the timing result executing only the RTA filtering randomly selected actions on all COTS processors. \tabref{tab:nnc_rta_cots} shows the timing result executing the NNC and RTA in series on all COTS processors.  

\begin{table}[H]
    \centering
    \caption{Neural Network Controller only (all sensors) for COTS processors (ms)}
    \label{tab:nnc_cots}
    % \resizebox{\linewidth}{!}{
    \begin{tabular}{lrrrrrr}
    \toprule
    Metric Labels & IQM & Mean & St. Dev. & Min & Median & Max \\
    \midrule
    AGX Orin (safe) & 1.187445 & 1.228211 & 0.609142 & 0.234604 & 1.152396 & 2.985716 \\
    AGX Orin (not safe) & 1.274283 & 1.251544 & 0.544837 & 0.260115 & 1.270294 & 3.172636 \\
    AGX Xavier (safe) & 0.955370 & 0.987614 & 0.116162 & 0.778198 & 0.937819 & 1.774073 \\
    AGX Xavier (not safe) & 0.943299 & 0.985999 & 0.218077 & 0.553131 & 0.849485 & 2.014160 \\
    Orin Nano (safe) & 0.826063 & 0.832814 & 0.116105 & 0.621319 & 0.826120 & 3.330231 \\
    Orin Nano (not safe) & 0.816813 & 0.821106 & 0.156956 & 0.595808 & 0.817776 & 5.327463 \\
    \bottomrule
    \end{tabular}
    % }
\end{table}

\begin{table}[H]
    \centering
    \caption{Run Time Assurance only for COTS processors (ms)}
    \label{tab:rta_cots}
    \resizebox{\linewidth}{!}{
    \begin{tabular}{lrrrrrr}
    \toprule
    Metric Labels & IQM & Mean & St. Dev. & Min & Median & Max \\
    \midrule
    AGX Orin eASIF (safe) & 2.750907 & 3.329385 & 1.815636 & 1.415491 & 2.644897 & 13.191223 \\
    AGX Orin eASIF (not safe) & 3.828257 & 4.387763 & 2.386426 & 1.437187 & 3.249407 & 14.296770 \\
    AGX Orin dASIF dt=1 (safe) & 16.464000 & 17.857730 & 7.505805 & 6.291151 & 16.286969 & 47.605753 \\
    AGX Orin dASIF dt=1 (not safe) & 13.954637 & 15.005995 & 6.395238 & 6.036758 & 13.627291 & 42.531967 \\
    AGX Orin dASIF dt=10 (safe) & 21.848526 & 21.802646 & 7.986108 & 6.202459 & 22.051692 & 50.238371 \\
    AGX Orin dASIF dt=10 (not safe) & 23.385702 & 23.700129 & 15.179291 & 5.964756 & 24.177551 & 271.273613 \\
    AGX Orin iASIF dt=1 (safe) & 12.569212 & 12.836350 & 2.253597 & 8.590221 & 12.634397 & 27.595997 \\
    AGX Orin iASIF dt=1 (not safe) & 12.351090 & 12.849332 & 2.830778 & 8.388996 & 12.449384 & 26.875496 \\
    AGX Orin iASIF dt=10 (safe) & 6.106794 & 6.150317 & 2.833092 & 1.900911 & 6.414652 & 19.070864 \\
    AGX Orin iASIF dt=10 (not safe) & 8.107551 & 8.085205 & 2.096096 & 1.858711 & 8.245826 & 16.801119 \\
    AGX Xavier eASIF (safe) & 6.780652 & 6.782235 & 1.046182 & 4.975796 & 6.831884 & 10.135412 \\
    AGX Xavier eASIF (not safe) & 6.897135 & 6.911544 & 1.385440 & 4.483938 & 6.967545 & 37.313461 \\
    AGX Xavier dASIF dt=1 (safe) & 21.084351 & 22.639788 & 8.528005 & 13.042927 & 22.226334 & 59.758186 \\
    AGX Xavier dASIF dt=1 (not safe) & 22.660113 & 24.019467 & 9.195153 & 13.480902 & 22.806048 & 66.515207 \\
    AGX Xavier dASIF dt=10 (safe) & 36.319591 & 33.975447 & 10.832311 & 13.076067 & 39.674759 & 99.483728 \\
    AGX Xavier dASIF dt=10 (not safe) & 35.776018 & 35.831206 & 30.146673 & 13.020277 & 38.898706 & 578.676701 \\
    AGX Xavier iASIF dt=1 (safe) & 21.970631 & 21.874224 & 1.098860 & 19.713163 & 21.958470 & 26.401520 \\
    AGX Xavier iASIF dt=1 (not safe) & 20.737039 & 21.076036 & 1.191656 & 19.755125 & 20.685315 & 38.733482 \\
    AGX Xavier iASIF dt=10 (safe) & 8.145295 & 8.161349 & 0.850160 & 6.367683 & 8.159876 & 10.876656 \\
    AGX Xavier iASIF dt=10 (not safe) & 35.422599 & 35.237457 & 27.231706 & 12.780905 & 38.780451 & 542.731047 \\
    Orin Nano eASIF (safe) & 4.355317 & 4.319885 & 0.285863 & 2.608538 & 4.352689 & 4.872799 \\
    Orin Nano eASIF (not safe) & 4.369097 & 4.489313 & 4.801894 & 2.299070 & 4.402399 & 147.660494 \\
    Orin Nano dASIF dt=1 (safe) & 15.896380 & 17.493387 & 8.630736 & 9.178877 & 16.353846 & 173.128605 \\
    Orin Nano dASIF dt=1 (not safe) & 17.178587 & 18.459619 & 7.733927 & 9.151459 & 17.144203 & 56.777000 \\
    Orin Nano dASIF dt=10 (safe) & 24.952983 & 23.636972 & 9.078023 & 9.112358 & 27.179956 & 173.389673 \\
    Orin Nano dASIF dt=10 (not safe) & 24.828254 & 24.594033 & 17.942731 & 9.175301 & 27.089834 & 366.775274 \\
    Orin Nano iASIF dt=1 (safe) & 17.348053 & 17.301687 & 1.035882 & 12.364388 & 17.362356 & 24.291277 \\
    Orin Nano iASIF dt=1 (not safe) & 17.102747 & 17.426064 & 2.161709 & 12.775660 & 17.038345 & 63.579559 \\
    Orin Nano iASIF dt=10 (safe) & 5.636376 & 5.587431 & 0.376810 & 3.245831 & 5.637646 & 6.979465 \\
    Orin Nano iASIF dt=10 (not safe) & 5.691443 & 5.645815 & 0.359121 & 3.118992 & 5.692244 & 7.040024 \\
    \bottomrule
    \end{tabular}
    }
\end{table}

\begin{table}[H]
    \centering
    \caption{Neural Network Controller (all sensors) and Run Time Assurance for COTS processors (ms)}
    \label{tab:nnc_rta_cots}
    \resizebox{\linewidth}{!}{
    \begin{tabular}{lrrrrrr}
    \toprule
    Metric Labels & IQM & Mean & St. Dev. & Min & Median & Max \\
    \midrule
    AGX Orin eASIF (safe) & 2.416291 & 2.554553 & 1.128687 & 1.048565 & 2.466440 & 8.322716 \\
    AGX Orin eASIF (not safe) & 4.610065 & 4.461648 & 1.998004 & 1.109600 & 5.039692 & 15.039921 \\
    AGX Orin dASIF dt=1 (safe) & 9.103857 & 9.555218 & 2.373391 & 5.664825 & 9.035110 & 26.257515 \\
    AGX Orin dASIF dt=1 (not safe) & 12.852021 & 13.881633 & 5.468681 & 5.748987 & 12.873411 & 44.796944 \\
    AGX Orin dASIF dt=10 (safe) & 9.476149 & 9.678036 & 1.897095 & 5.804539 & 9.476662 & 16.189337 \\
    AGX Orin dASIF dt=10 (not safe) & 10.646212 & 13.780409 & 30.656480 & 5.413771 & 10.231495 & 886.292696 \\
    AGX Orin iASIF dt=1 (safe) & 13.380730 & 13.704895 & 3.052360 & 8.404255 & 13.133049 & 28.942347 \\
    AGX Orin iASIF dt=1 (not safe) & 13.348919 & 13.793151 & 3.104095 & 8.520126 & 13.067245 & 31.623125 \\
    AGX Orin iASIF dt=10 (safe) & 3.296854 & 3.630646 & 1.704423 & 1.539707 & 3.324151 & 13.426065 \\
    AGX Orin iASIF dt=10 (not safe) & 5.201894 & 5.252436 & 2.341010 & 1.563787 & 5.368471 & 20.745516 \\
    AGX Xavier eASIF (safe) & 4.372336 & 4.395119 & 0.579755 & 2.779245 & 4.352331 & 6.491184 \\
    AGX Xavier eASIF (not safe) & 4.515870 & 4.493873 & 0.653715 & 2.696514 & 4.544854 & 6.919384 \\
    AGX Xavier dASIF dt=1 (safe) & 15.325570 & 15.647139 & 2.177019 & 12.193918 & 15.392542 & 26.516199 \\
    AGX Xavier dASIF dt=1 (not safe) & 14.981482 & 16.886219 & 5.844275 & 12.039185 & 15.101910 & 49.370050 \\
    AGX Xavier dASIF dt=10 (safe) & 14.652752 & 15.066409 & 2.398740 & 11.744261 & 14.610052 & 28.725147 \\
    AGX Xavier dASIF dt=10 (not safe) & 14.623624 & 20.185279 & 65.600733 & 11.199236 & 14.662147 & 1925.134420 \\
    AGX Xavier iASIF dt=1 (safe) & 20.191029 & 20.298672 & 1.923566 & 18.255949 & 20.744324 & 45.238972 \\
    AGX Xavier iASIF dt=1 (not safe) & 19.351970 & 19.717433 & 1.221049 & 18.252373 & 19.337416 & 25.688648 \\
    AGX Xavier iASIF dt=10 (safe) & 5.933733 & 5.912681 & 0.761662 & 4.300117 & 5.967855 & 8.225441 \\
    AGX Xavier iASIF dt=10 (not safe) & 6.097788 & 6.139364 & 0.554479 & 3.811359 & 6.030202 & 8.520126 \\
    \bottomrule
    \end{tabular}
    }
\end{table}

\subsection{Radiation Tolerant Timing Results}

The full timing results from running the experiments on radiation tolerant hardware are shown in the following tables. These results come from measuring the execution time for 10,000 test cases for each row. \tabref{tab:nnc_ix10} shows the timing result executing only the NNC on the radiation tolerant processor. \tabref{tab:rta_ix10} shows the timing result executing only the RTA filtering randomly selected actions on the radiation tolerant processor. \tabref{tab:nnc_rta_ix10} shows the timing result executing the NNC and RTA in series on the radiation tolerant processor. 

\begin{table}[H]
    \centering
    \caption{Neural Network Controller only for Unibap IX10 (ms)}
    \label{tab:nnc_ix10}
    \resizebox{\linewidth}{!}{
    \begin{tabular}{lrrrrrrr}
    \toprule
    Metric Labels & IQM & Mean & St. Dev. & Min & Median & Max & Execution with JIT \\
    \midrule
    all sensors (safe) & 0.024680 & 0.025994 & 0.004672 & 0.021951 & 0.024636 & 0.114974 & 0.790562 \\
    all sensors (not safe) & 0.025528 & 0.026237 & 0.003622 & 0.022622 & 0.024786 & 0.156452 & 0.120034 \\
    no sensors (safe) & 0.023983 & 0.025586 & 0.004621 & 0.022382 & 0.023504 & 0.080120 & 0.358081 \\
    no sensors (not safe) & 0.024736 & 0.025361 & 0.003284 & 0.021911 & 0.024566 & 0.113533 & 0.316112 \\
    \bottomrule
    \end{tabular}
    }
\end{table}

\begin{table}[H]
    \centering
    \caption{Run Time Assurance only for Unibap IX10 (ms)}
    \label{tab:rta_ix10}
    \resizebox{\linewidth}{!}{
    \begin{tabular}{lrrrrrrr}
    \toprule
    Metric Labels & IQM & Mean & St. Dev. & Min & Median & Max \\
    \midrule
    eASIF (safe) & 0.117673 & 0.123958 & 0.018850 & 0.107168 & 0.117154 & 0.403735 \\
    eASIF (not safe) & 0.118868 & 0.122073 & 0.013497 & 0.107031 & 0.113589 & 0.421892 \\
    dASIF dt=1 tol=1e-3 (safe) & 4.214526 & 38.633448 & 584.449966 & 0.829635 & 3.944832 & 17361.761425 \\
    dASIF dt=1 tol=1e-3 (not safe) & 3.851534 & 11.168847 & 62.547584 & 0.793099 & 3.655996 & 1524.341135 \\
    dASIF dt=10 tol=1e-3 (safe) & 8.827194 & 317.120068 & 3624.942297 & 0.798718 & 8.441662 & 60000.189866 \\
    dASIF dt=10 tol=1e-3 (not safe) & 7.919051 & 54.155606 & 332.602006 & 0.836692 & 7.557296 & 7233.064712 \\
    dASIF dt=1 tol=1e-4 (safe) & 9.419962 & 154.741459 & 1872.711906 & 4.278250 & 9.677567 & 47860.026079 \\
    dASIF dt=1 tol=1e-4 (not safe) & 8.010065 & 55.016445 & 477.519387 & 4.194320 & 8.135115 & 9989.882471 \\
    dASIF dt=10 tol=1e-4 (safe) & 14.932394 & 448.032044 & 4123.310098 & 4.202426 & 13.971648 & 60000.149681 \\
    dASIF dt=10 tol=1e-4 (not safe) & 12.901447 & 371.274909 & 3696.299910 & 4.190733 & 12.136475 & 60000.098485 \\
    iASIF dt=1 (safe) & 5.596083 & 5.573943 & 0.600580 & 4.672187 & 5.607997 & 8.721112 \\
    iASIF dt=1 (not safe) & 5.421894 & 5.439027 & 0.609651 & 4.637850 & 5.294439 & 9.210239 \\
    iASIF dt=10 (safe) & 0.546723 & 0.554855 & 0.062099 & 0.478004 & 0.513545 & 0.788635 \\
    iASIF dt=10 (not safe) & 0.494691 & 0.505387 & 0.039198 & 0.471982 & 0.494978 & 0.806314 \\
    \bottomrule
    \end{tabular}
    }
\end{table}

\begin{table}[H]
    \centering
    \caption{Neural Network Controller and Run Time Assurance  for Unibap IX10 (ms)}
    \label{tab:nnc_rta_ix10}
    \resizebox{\linewidth}{!}{
    \begin{tabular}{lrrrrrrr}
    \toprule
    Metric Labels & IQM & Mean & St. Dev. & Min & Median & Max & Execution with JIT \\
    \midrule
    eASIF all sensors (safe) & 0.144337 & 0.148043 & 0.021234 & 0.140416 & 0.144263 & 1.998211 & 0.320026 \\
    eASIF all sensors (not safe) & 0.145421 & 0.152326 & 0.016319 & 0.140015 & 0.144403 & 0.283246 & 0.284448 \\
    eASIF no sensors (safe) & 0.148515 & 0.155292 & 0.018306 & 0.139888 & 0.148204 & 0.356702 & 0.347584 \\
    eASIF no sensors (not safe) & 0.148306 & 0.153700 & 0.016972 & 0.139216 & 0.148113 & 0.395676 & 0.320242 \\
    dASIF dt=1 tol=1e-3 all sensors (safe) & 3.721660 & 4.058949 & 1.240831 & 1.409485 & 3.714212 & 11.769172 & 6.820636 \\
    dASIF dt=1 tol=1e-3 all sensors (not safe) & 3.708151 & 4.061183 & 1.271563 & 1.151727 & 3.625291 & 11.576846 & 5.041455 \\
    dASIF dt=1 tol=1e-3 no sensors (safe) & 3.820180 & 11.449278 & 70.853788 & 0.854356 & 3.786309 & 1263.299591 & 3.359823 \\
    dASIF dt=1 tol=1e-3 no sensors (not safe) & 3.938696 & 44.312923 & 1100.413853 & 0.850408 & 3.806449 & 34762.922707 & 2.928989 \\
    dASIF dt=10 tol=1e-3 all sensors (safe) & 4.120237 & 4.509934 & 1.530665 & 1.434351 & 3.950538 & 13.016087 & 9.242374 \\
    dASIF dt=10 tol=1e-3 all sensors (not safe) & 4.203840 & 4.570982 & 1.591459 & 1.429740 & 4.030370 & 30.398838 & 9.479227 \\
    dASIF dt=10 tol=1e-3 no sensors (safe) & 7.107587 & 68.480251 & 951.044295 & 0.876166 & 6.877964 & 28919.426470 & 3.655862 \\
    dASIF dt=10 tol=1e-3 no sensors (not safe) & 6.722149 & 47.481896 & 422.050374 & 0.867370 & 6.573830 & 10118.682149 & 4.680501 \\
    dASIF dt=1 tol=1e-4 all sensors (safe) & 6.075389 & 6.691481 & 1.847697 & 4.435233 & 5.986329 & 16.914064 & 12.874864 \\
    dASIF dt=1 tol=1e-4 all sensors (not safe) & 6.081020 & 6.656225 & 1.853271 & 4.421535 & 5.899599 & 19.748987 & 8.555593 \\
    dASIF dt=1 tol=1e-4 no sensors (safe) & 8.719481 & 85.466763 & 1900.920043 & 2.619888 & 8.856243 & 60000.210210 & 8.787915 \\
    dASIF dt=1 tol=1e-4 no sensors (not safe) & 8.403860 & 118.391181 & 1610.316620 & 2.576704 & 8.586583 & 32876.944156 & 3.994396 \\
    dASIF dt=10 tol=1e-4 all sensors (safe) & 7.035054 & 7.514595 & 2.371262 & 4.521534 & 6.301900 & 20.624789 & 14.882400 \\
    dASIF dt=10 tol=1e-4 all sensors (not safe) & 7.170553 & 7.600403 & 2.567077 & 4.440335 & 6.454838 & 62.132830 & 14.325055 \\
    dASIF dt=10 tol=1e-4 no sensors (safe) & 12.820199 & 219.127918 & 2765.933881 & 2.769523 & 12.631502 & 60000.123889 & 12.427414 \\
    dASIF dt=10 tol=1e-4 no sensors (not safe) & 11.966611 & 331.040660 & 3731.542520 & 2.683378 & 11.765680 & 60000.235233 & 15.292397 \\
    iASIF dt=1 all sensors (safe) & 5.394485 & 5.523652 & 0.626012 & 4.872785 & 5.198343 & 9.845719 & 7.048852 \\
    iASIF dt=1 all sensors (not safe) & 5.397487 & 5.534790 & 0.604243 & 4.909055 & 5.242497 & 8.479896 & 8.634139 \\
    iASIF dt=1 no sensors (safe) & 5.216871 & 5.421575 & 0.551656 & 4.912907 & 5.103218 & 9.782077 & 6.681380 \\
    iASIF dt=1 no sensors (not safe) & 5.578445 & 5.634629 & 0.593101 & 4.950629 & 5.461640 & 8.955738 & 9.248657 \\
    iASIF dt=10 all sensors (safe) & 0.572849 & 0.588172 & 0.070050 & 0.524144 & 0.536647 & 0.948226 & 1.069346 \\
    iASIF dt=10 all sensors (not safe) & 0.536340 & 0.560853 & 0.065782 & 0.523822 & 0.534903 & 1.224569 & 0.795307 \\
    iASIF dt=10 no sensors (safe) & 0.547699 & 0.576585 & 0.057893 & 0.533684 & 0.545427 & 0.825248 & 0.799219 \\
    iASIF dt=10 no sensors (not safe) & 0.590823 & 0.597209 & 0.066921 & 0.525249 & 0.548333 & 0.976318 & 0.819318 \\
    \bottomrule
    \end{tabular}
    }
\end{table}

\section{Discussion} \label{sec:discussion}

This section discusses the main results and trends found while running experiments on each processor. The full results are presented in \secref{sec:results}. 

\subsection{Commercial Off The Shelf NVIDIA Boards}

\begin{figure}[htbp]
    \centering
    \includegraphics[width=\columnwidth]{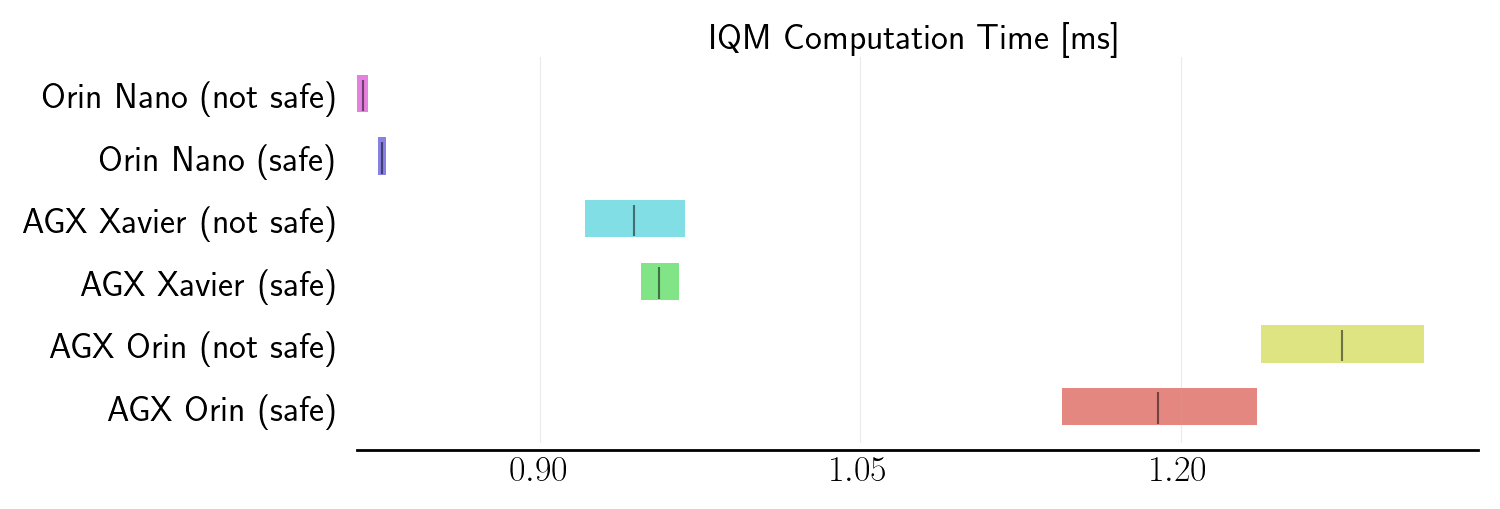}
    \caption{NNC only (all sensors)}
    \label{fig:cots_nnc_only}
\end{figure}

\figref{fig:cots_nnc_only} shows the IQM of computation time for the NNC only across each COTS processor. The data shows that the Orin Nano is able to run the NNC the fastest, while the AGX Orin is the slowest. In this case, safe states and not safe states have very little difference on the computation time of the NNC. On any processor, the NNC is typically able to run in less than 1.5~ms, and \tabref{tab:nnc_cots} shows the MOET is 5.3~ms.

\begin{figure}[htbp]
    \centering
    \includegraphics[width=\columnwidth]{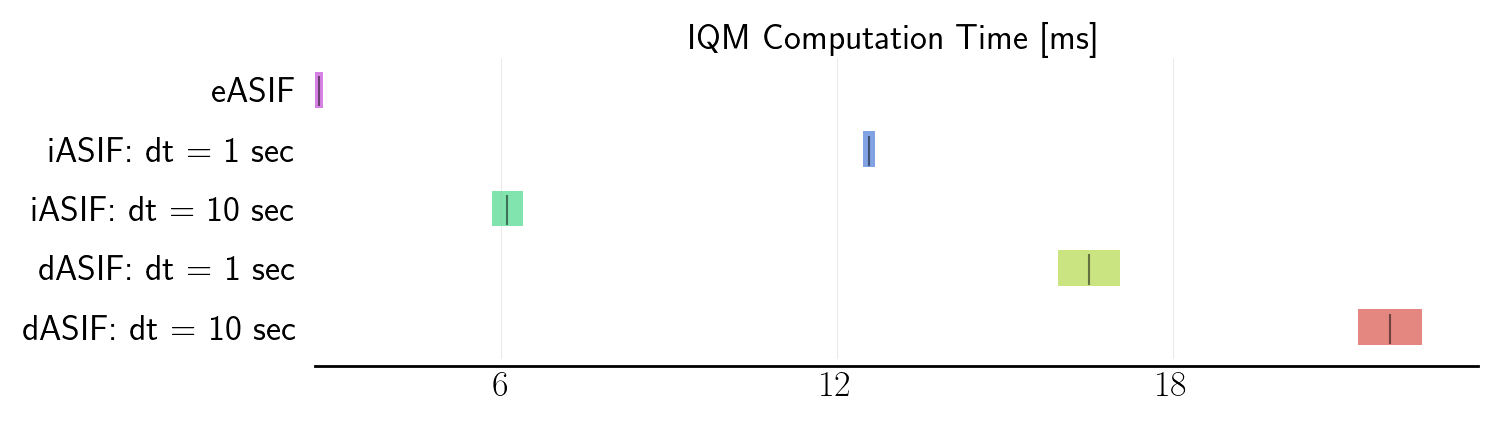}
    \caption{RTA only - AGX Orin}
    \label{fig:agx_orin_rta}
\end{figure}

\figref{fig:agx_orin_rta} shows the computation time for each RTA configuration on the AGX Orin, all with safe states only. While only one processor is shown, the results show a similar trend found on the other processors. eASIF is shown to be the fastest configuration at less than 5~ms, which is intuitive as it requires the least computation to construct the barrier constraints. iASIF is the next fastest, where a 10~s timestep is faster than a 1~s timestep because it requires fewer trajectory states to be computed. dASIF is shown to be the slowest, notably because of the nonlinear solver. A 1~s timstep is faster than a 10~s timestep in this case because a state 1~s into the future is more likely to be safe than a state 10~s into the future, and safe states allow the nonlinear solver to find a safe solution faster. \tabref{tab:rta_cots} shows that the MOET for eASIF, iASIF, and dASIF using safe states are 13.2, 27.6, and 173.4~ms respectively. The MOET for dASIF is clearly much slower, which again enforces that nonlinear solvers are incomplete and can cause long computation times. While eASIF and iASIF both have advantages to the designer, in most cases dASIF should only be used over eASIF if it cannot run fast enough in real time for continuous control.

\begin{figure}[htbp]
    \centering
    \includegraphics[width=\columnwidth]{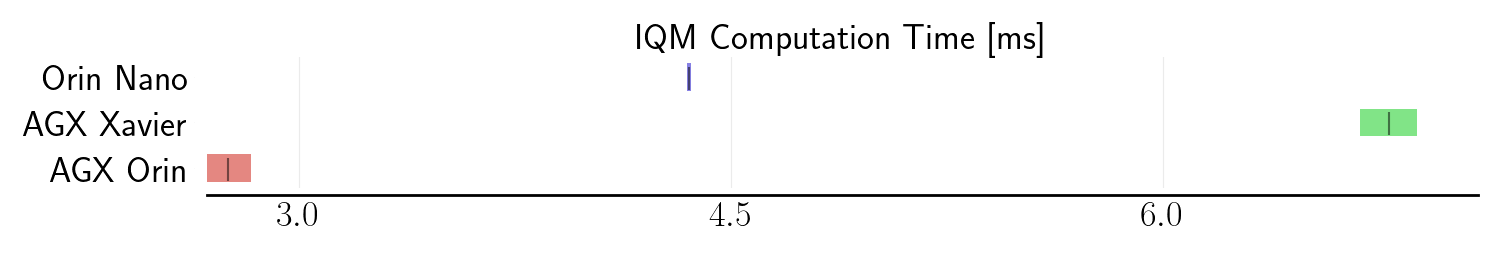}
    \caption{RTA only (eASIF) - comparing processors}
    \label{fig:eASIF_rta_only}
\end{figure}

\figref{fig:eASIF_rta_only} shows the computation time for eASIF on each processor. The data shows that the AGX Orin is the fastest in this case, while the AGX Xavier is the slowest. This trend is similar for all of the RTA configurations. These results show that different processors provide different advantages for running the NNC or RTA.

\begin{figure}[htbp]
    \centering
    \includegraphics[width=\columnwidth]{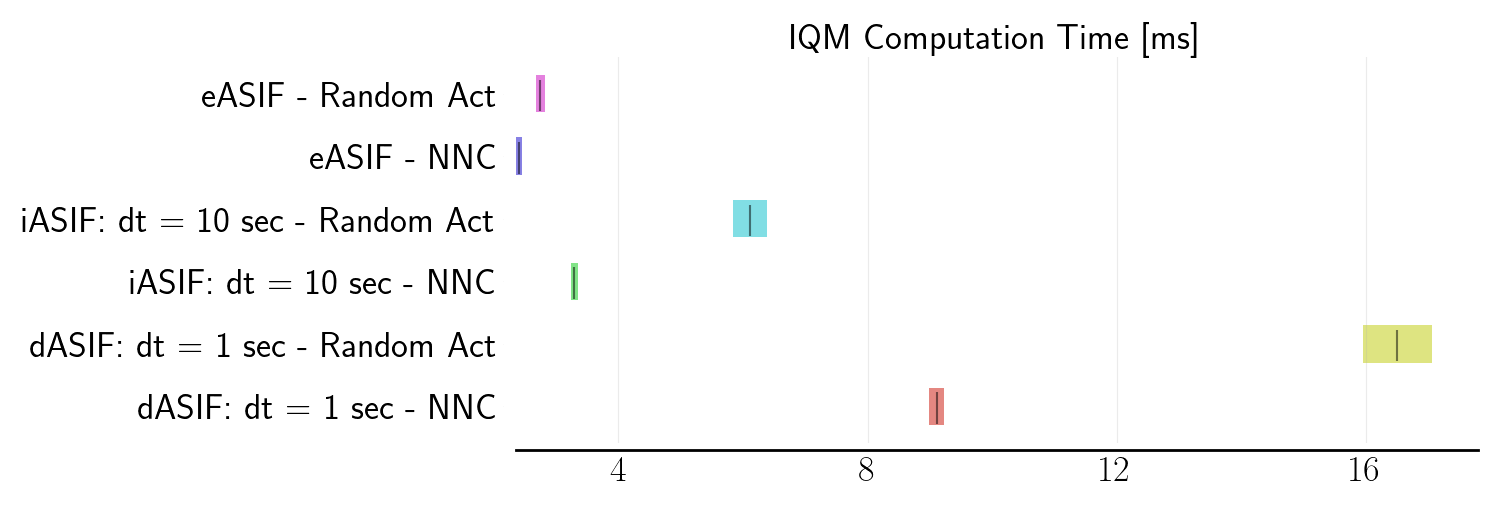}
    \caption{RTA + NNC - AGX Orin}
    \label{fig:agx_orin_rta_nnc}
\end{figure}

Finally, \figref{fig:agx_orin_rta_nnc} shows the computation time for the NNC combined with RTA for the AGX Orin, compared to RTA only with random actions. The best performing RTA configurations are shown, but the trends are similar for all other configurations and processors. The data shows that for each RTA configuration, combining it with the NNC is faster than using random actions. This is because the NNC is more likely to produce safe actions, which allows the RTA to find a safe solution faster. This also shows that the computation time for the NNC combined with the RTA is not simply adding the computation time of the NNC only and RTA only, but rather the two systems can work together to reduce the total time.

\subsection{Radiation Tolerant Unibap iX10}

\begin{figure}[htbp]
    \centering
    \includegraphics[width=\columnwidth]{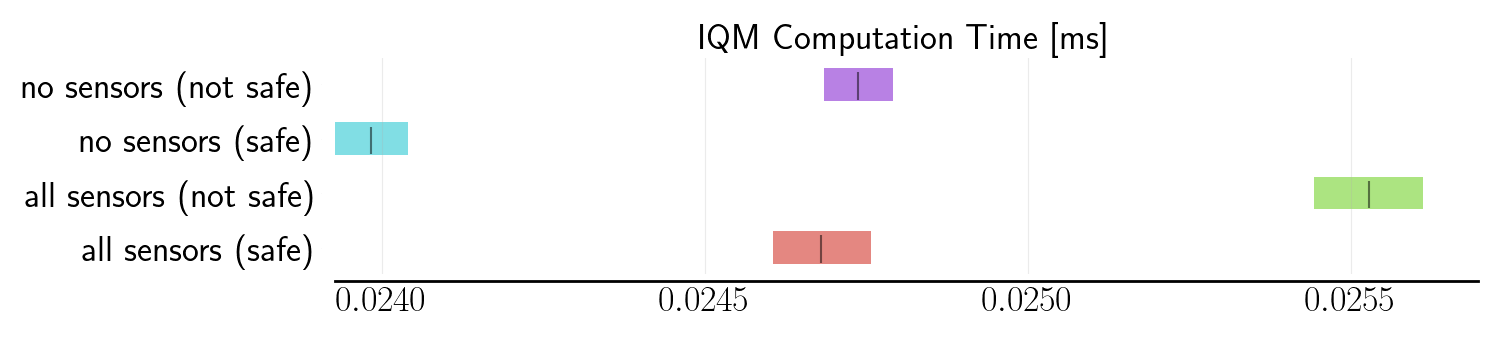}
    \caption{NNC only}
    \label{fig:nnc_only}
\end{figure}

\figref{fig:nnc_only} shows the computation time for NNC only on the iX10. When comparing the two different NNCs, the data shows that the no sensors configuration is faster than the all sensors configuration because it has less model inputs and requires less computation. It can also be seen that safe states are faster than not safe states, because they are less likely to match the scenarios used to train the NNC. Overall, the NNC is typically able to run in less than 0.03~ms, and \tabref{tab:nnc_ix10} shows the MOET is 0.16~ms. 

\begin{figure}[htbp]
    \centering
    \includegraphics[width=\columnwidth]{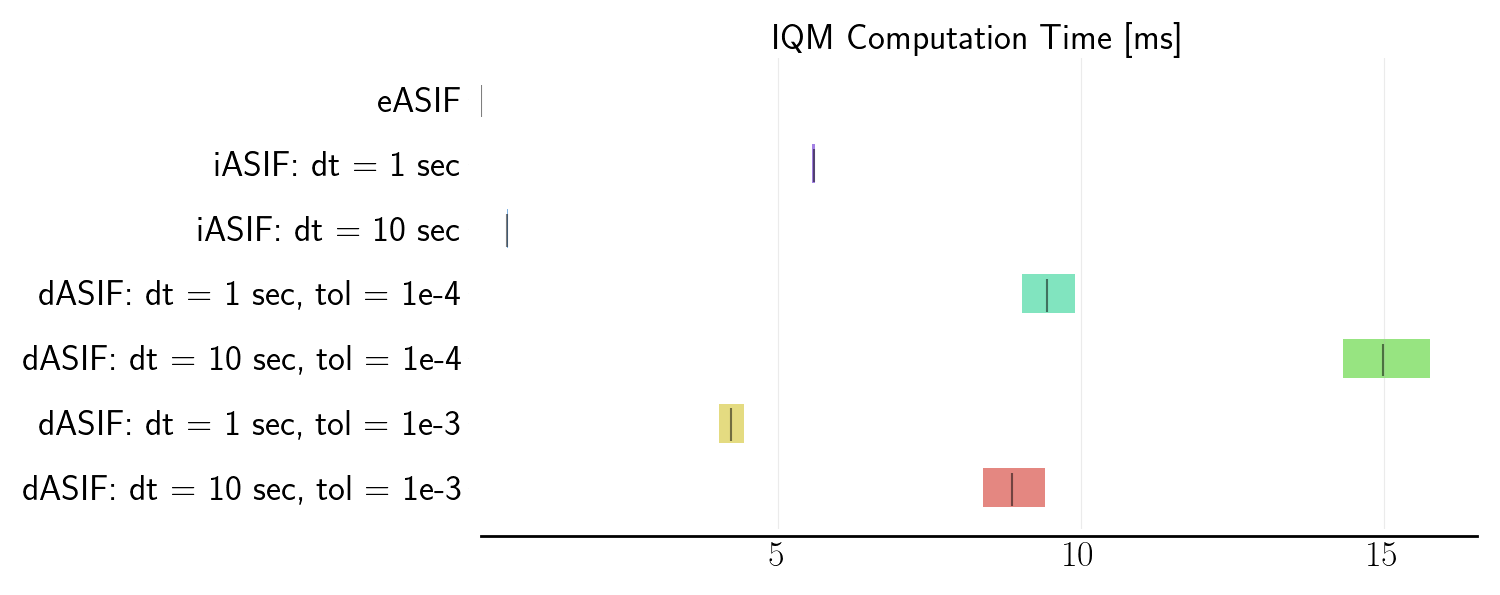}
    \caption{RTA only}
    \label{fig:rta_only}
\end{figure}

\figref{fig:rta_only} shows the computation time for RTA only, with safe states only. The results follow the trends of \figref{fig:agx_orin_rta}, where eASIF is the fastest and dASIF is the slowest. For dASIF, having a smaller tolerance requires more time for the solver to find a solution because it must be more precise. \tabref{tab:rta_ix10} shows that the MOET for eASIF, iASIF, and dASIF using safe states are 0.40, 8.72, and 60,000~ms respectively, where the nonlinear solver was stopped at 60~s. Note that the dASIF times could be greatly reduced with further optimization to the nonlinear solver and potentially selecting a different algorithm to solve the problem.

\begin{figure}[htbp]
    \centering
    \includegraphics[width=\columnwidth]{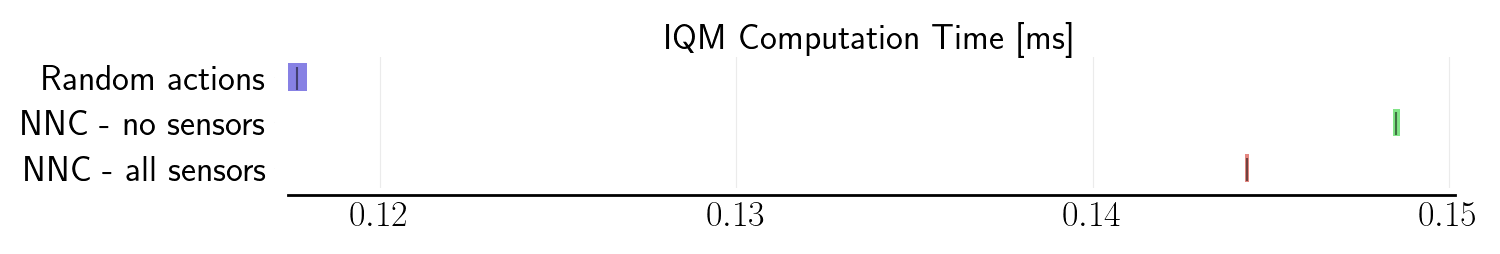}
    \caption{NNC + eASIF}
    \label{fig:nnc_rta_eASIF}
\end{figure}

\begin{figure}[htbp]
    \centering
    \includegraphics[width=\columnwidth]{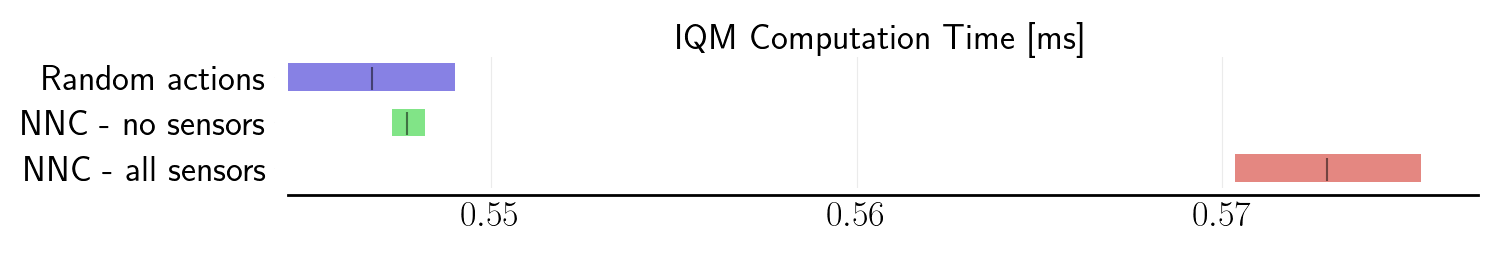}
    \caption{NNC + iASIF: dt = 10}
    \label{fig:nnc_rta_iASIF}
\end{figure}

\begin{figure}[htbp]
    \centering
    \includegraphics[width=\columnwidth]{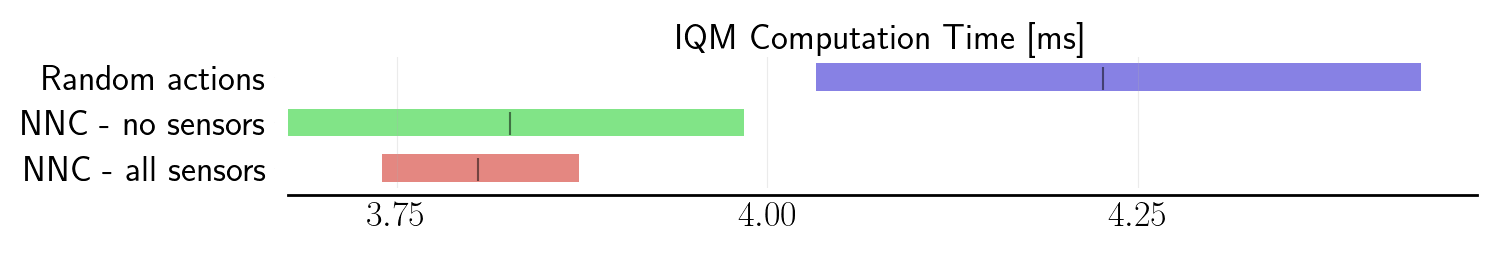}
    \caption{NNC + dASIF: dt = 1, tol = 1e-3}
    \label{fig:nnc_rta_dASIF}
\end{figure}

\figref{fig:nnc_rta_eASIF} shows the computation time for eASIF only compared to the NNC combined with eASIF, and \figref{fig:nnc_rta_iASIF} and \figref{fig:nnc_rta_dASIF} show the same comparison for the best performing iASIF and dASIF configurations respectively. For eASIF and iASIF, the data shows that using random actions is faster than combining the NNC and RTA because the computation time for matrix conversion outweighs the computation time for safer actions. However, for dASIF, safe actions are more impactful in lowering the computation time as they allow the nonlinear solver to compute the correct solution in fewer steps.

%%%%%%%%%%%%%%%%%%%%%%%%%%%%%%%%%%%%%%%%%%%%%%%%%%%%%%%%%%%%%%%%%%%%%%
% Conclusion
%%%%%%%%%%%%%%%%%%%%%%%%%%%%%%%%%%%%%%%%%%%%%%%%%%%%%%%%%%%%%%%%%%%%%%
\section{Conclusion}

This paper presented many different experiments where an NNC and RTA filter were run on COTS and radiation tolerant processors. 
% Based on the MOET for each each setup, the NNC alone can be run at a rate of 200~Hz on a COTS processor and 3000~Hz on a radiation tolerant processor. 
For the spacecraft inspection task defined in this paper, it is estimated that a control system consisting of an NNC and RTA filter would not need to run faster than 10~Hz in real time. Based on the MOET for each configuration tested, this can be achieved for all configurations on COTS and radiation tolerant processors except for dASIF. It is recommended that eASIF be used on any of these processors over dASIF since it has the ability to run in real time. However, the dASIF algorithm could be further optimized by adjusting solver parameters such as the algorithm used. In general, all algorithms tested in this paper should be further optimized before application to a real world platform. Importantly, it was found that using a primary controller that produces safe actions allows an RTA filter to solve the optimization problem faster.
While many other algorithms and processes must be run on board space grade hardware, these results suggest that an NNC and RTA filter can be run in real time on board COTS and radiation tolerant processors. Future work will consist of repeating these experiments on radiation hardened processors.

% \section*{Appendix}

\section*{Acknowledgments}
This research was sponsored by the Air Force Research Laboratory under the \textit{Safe Trusted Autonomy for Responsible Spacecraft} (STARS) Seedlings for Disruptive Capabilities Program.
The views expressed are those of the authors and do not reflect the official guidance or position of the United States Government, the Department of Defense, or of the United States Air Force. This work has been approved for public release: distribution unlimited. Case Number AFRL-2024-2470.

\bibliography{references}

\end{document}